\documentclass[twocolumn,secnumarabic,superscriptaddress,amssymb,nobibnotes,prapplied]{revtex4-2}
\usepackage{graphicx}
\usepackage{dcolumn}
\usepackage{bm}
\usepackage{natbib}
\usepackage{floatrow}
\usepackage{subfig}
\usepackage{hyperref}
\hypersetup{colorlinks,allcolors=blue}

\begin{document}

\title{nanoTesla magnetometry with the silicon vacancy in silicon carbide}

\author{John B. S. Abraham}
\email{John.Abraham@jhuapl.edu}
\affiliation{Johns Hopkins University Applied Physics Laboratory, Laurel, MD, 20723, USA}
\author{Cameron Gutgsell}
\affiliation{Johns Hopkins University Applied Physics Laboratory, Laurel, MD, 20723, USA}
\author{Dalibor Todorovski}
\affiliation{Johns Hopkins University Applied Physics Laboratory, Laurel, MD, 20723, USA}
\author{Scott Sperling}
\altaffiliation[Present address: ]{Northrop Grumman, Linthicum Heights, MD, 21090, USA}
\affiliation{Johns Hopkins University Applied Physics Laboratory, Laurel, MD, 20723, USA}
\author{Jacob E. Epstein}
\altaffiliation[Present address: ]{Starry, Boston, MA, 02111, USA}
\affiliation{Johns Hopkins University Applied Physics Laboratory, Laurel, MD, 20723, USA}
\author{Brian S. Tien-Street}
\affiliation{Johns Hopkins University Applied Physics Laboratory, Laurel, MD, 20723, USA}
\author{Timothy M. Sweeney}
\affiliation{Johns Hopkins University Applied Physics Laboratory, Laurel, MD, 20723, USA}
\author{Jeremiah J. Wathen}
\affiliation{Johns Hopkins University Applied Physics Laboratory, Laurel, MD, 20723, USA}
\author{Elizabeth A. Pogue}
\affiliation{Department of Chemistry, Johns Hopkins University, Baltimore, MD, 21218, USA}
\affiliation{Institute for Quantum Matter, Department of Physics and Astronomy, Johns Hopkins University, Baltimore, MD, 21218, USA}
\author{Peter G. Brereton}
\affiliation{Department of Physics, United States Naval Academy, Annapolis, MD, 21402, USA }
\author{Tyrel M. McQueen}
\affiliation{Department of Chemistry, Johns Hopkins University, Baltimore, MD, 21218, USA}
\affiliation{Institute for Quantum Matter, Department of Physics and Astronomy, Johns Hopkins University, Baltimore, MD, 21218, USA}
\affiliation{Department of Materials Science and Engineering, Johns Hopkins University, Baltimore, MD, 21218, USA}
\author{Wesley Frey}
\affiliation{McClellan Nuclear Research Center - UC Davis, McClellan, CA, 95652, USA}
\author{B. D. Clader}
\affiliation{Johns Hopkins University Applied Physics Laboratory, Laurel, MD, 20723, USA}
\author{Robert Osiander}
\affiliation{Johns Hopkins University Applied Physics Laboratory, Laurel, MD, 20723, USA}

\date{\today}

\begin{abstract}
Silicon Carbide is a promising host material for spin defect based quantum sensors owing to its commercial availability and established techniques for electrical and optical microfabricated device integration. The negatively charged silicon vacancy is one of the leading spin defects studied in silicon carbide owing to its near telecom photoemission, high spin number, and nearly temperature independent ground state zero field splitting. We report the realization of nanoTesla shot-noise limited ensemble magnetometry based on optically detected magnetic resonance with the silicon vacancy in 4H silicon carbide. By coarsely optimizing the anneal parameters and minimizing power broadening, we achieved a sensitivity of 50 nT/$\sqrt{Hz}$ and a theoretical shot-noise-limited sensitivity of 3.5 nT/$\sqrt{Hz}$. This was accomplished without utilizing complex photonic engineering, control protocols, or applying excitation powers greater than a Watt. This work demonstrates that the silicon vacancy in silicon carbide provides a low-cost and simple approach to quantum sensing of magnetic fields.
\end{abstract}

\maketitle

\section{Introduction}
Spin defect quantum sensing \cite{RevModPhys.89.035002} holds promise to revolutionize a significant number of fields ranging from fundamental physics \cite{norman2020,atature2018} to biological systems \cite{barry2016,wu2016,price2020} due to their high sensitivity and spatial selectivity. Magnetic field sensing with spin defects is of significant interest \cite{rondin2014,atature2018,castelletto2020,barry2020}.  Recently, the negatively charged silicon mono-vacancy ($V^{-}_{Si}$) in silicon carbide 4H (SiC) has emerged as a promising spin defect for magnetometry and other applications \cite{tarasenko2018,castelletto2020} owing to a number of properties common with the diamond nitrogen vacancy (DNV) and other properties which are unique to the defect and host material. Defects can be generated in silicon carbide through through irradiation \cite{kasper2020} or thermal treatment \cite{vainer1981}. 

Like the DNV, the energy structure of the $V^{-}_{Si}$ has an inter-system crossing which spin polarizes the defect with optical pumping \cite{baranov2011,widmann2015} with coherence times comparable to those of DNV centers \cite{rondin2014,simin2017,kasper2020}. The spin of the $V^{-}_{Si}$ is 3/2. Thus, Kramer's degeneracy theorem applies to the symmetry of the defect ground state. This results in an essentially temperature independent ground state zero field splitting \cite{Anisimov2016}. Since the defect has a spin greater than one, multiple transitions can be interrogated \cite{neithammer2016,soltamov2019}. This has been utilized for vector magnetomtery \cite{neithammer2016}, all optical DC magnetometry \cite{simin2016}, and qudit (quantum states of dimension 4) magnetometry \cite{soltamov2019}. In addition to the compelling spin properties of the defect, it has been demonstrated that the $V^{-}_{Si}$ is able to be excited electrically and its state measured through spin dependent recombination \cite{lohrmann2015,cochrane2016}. This opens the possibility of operating $V^{-}_{Si}$ based quantum systems in an entirely electrical manner, which is an exciting proposition given the maturity of silicon carbide integrated device fabrication. 

\begin{figure*}[tp]	
\includegraphics[width=0.4\textwidth, height=0.23\textheight]{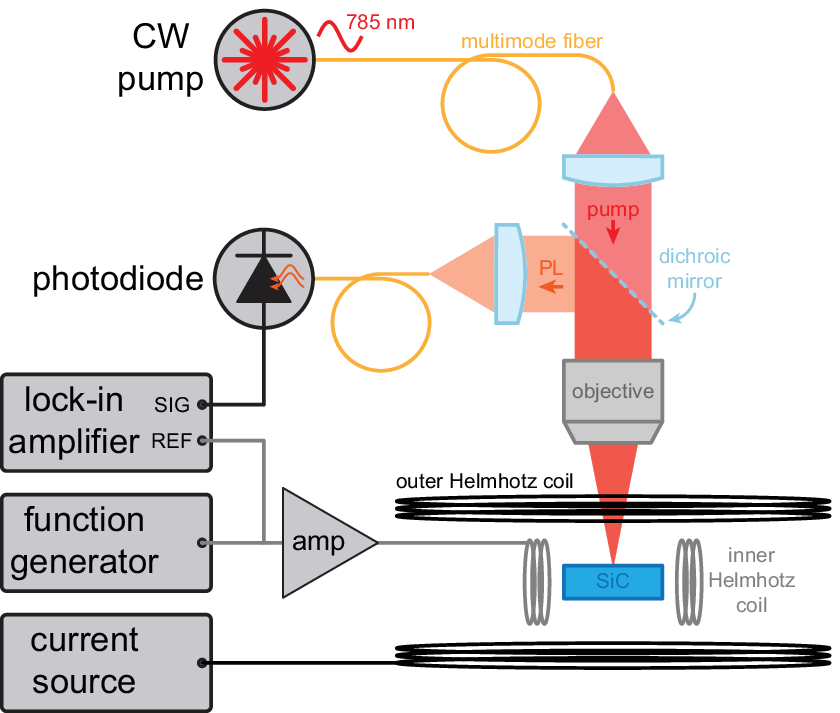}\label{exp}
\includegraphics[width=0.4\textwidth, height=0.23\textheight]{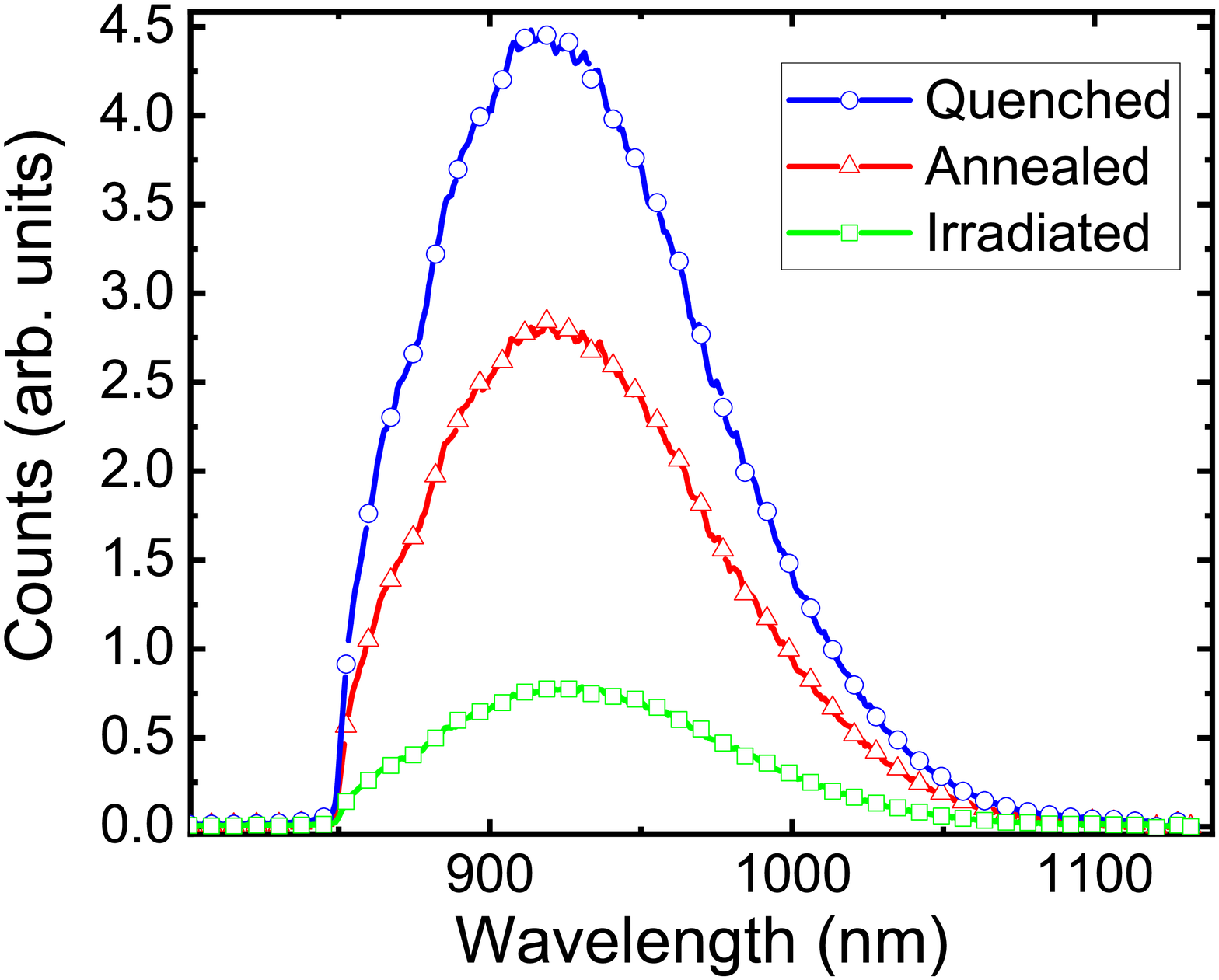}\label{pl}
\caption{(a) The Experimental setup for ODMR and power broadening measurements. A continuous-wave 785-nm laser is focused on the sample with a microscope objective. The sample fluorescence is collected by the same objective and is passed through a 850 nm long-pass filter and measured with a photodetector which is connected to the signal-in of a lock-in amplifier. The outer Helmholtz coil is driven by a current source to apply a static field along the c axis of the sample. The inner Helmholtz coil is driven by the function generator with a modulated RF tone. The function generator also outputs the modulation tone to the lock-in amplifier's reference channel. (b) Representative relative photoluminescence emission from the samples after irradiation, annealing, and thermal quenching following the anneal}\label{exp_pl}
\end{figure*}

Magnetometry with solid state spin defects has been demonstrated through a number of techniques with the DNV \cite{budker2017,barry2020}. Continuous wave optically detected magnetic resonance (CW ODMR) is one of the fundamental magnetometry techniques with such defects. Although it is not the most sensitive technique, the basic implementation is straight forward and is thus a good technique to baseline the relative sensitivity of different spin defects. The technique relies on the change in fluorescence intensity when an applied radio-frequency (RF) field is resonant with a ground state spin transition, typically of magnetic quantum number $m = 1$. The ground state transition energies are a function of the ambient magnetic field through the Zeeman effect. Thus, the ambient magnetic field is measured through the ground state transition energy or applied RF frequency through a change in the photoluminescence intensity \cite{shin2012,kraus2014-1,barry2016,simin2016,neithammer2016,clevenson2015,clevenson2018}. The ground state transition frequency is given by 

\begin{equation}
\nu = \nu_{ZFS} \pm {\gamma}B_{0}
\label{f_B}
\end{equation}

where $\nu_{ZFS}$ = 70 MHz is the zero field splitting for the $V^{-}_{Si}$ in 4H-SiC, $\gamma$ is the electron gyromagnetic ratio, and B$_{0}$ is the ambient magnetic field strength. The sensitivity of ODMR magnetometry is typically limited by the shot noise sensitivity, which stems from the statistics of photon emission from the spin defects \cite{shin2012,jensen2013,budker2017,clevenson2015,clevenson2018,wang2020}. The shot noise limited sensitivity has the following form \footnote{c.f. Eq. (8) of Ref. \cite{shin2012}. Our result varies by a factor of $\sqrt{2/3}$ from that result since we are not making a gradiometric measurement and we are reporting full-width half-max of our line-shape.}:

\begin{equation}
\eta_B = \frac{4\sqrt{2}}{3\sqrt{3}}\frac{h}{g\mu_B}\frac{\Delta}{C\sqrt{R}},
\label{sn_sensitivity}
\end{equation}
where $\Delta$ is the full width half maximum (FWHM) of the ODMR line-width, $C$ is the ODMR fluorescence contrast, $R$ is the rate of detected photons from the defect ensemble, $g\approx2.0032$ is the Land\'{e} g factor for the $V^{-}_{Si}$  spin defect \cite{soltamov2012}, and $h$ and $\mu_B$ are Planck's constant and the Bohr magneton respectively. The linewidth $\Delta$ is inversely proportional to the spin dephasing time of the defect, $T_{2}^{*}$, while $C$ and $R$ depend on many terms such as the Rabi frequency, transition rates, defect brightness, and optical collection efficiency.

In this article, we demonstrate an order of magnitude improvement in the sensitivity of CW ODMR magnetometry with an ensemble of $V^{-}_{Si}$ by optimizing the annealing parameters. We are able to achieve a shot noise sensitivity of 3.5 nT/$\sqrt{Hz}$. This is a significant improvement relative to previously published results \cite{simin2016, neithammer2016, wang2020}. We begin with a description of our sample preparation and the experimental set-up. From there we present our techniques and results.

\begin{figure*}
	\includegraphics[width=0.45\textwidth]{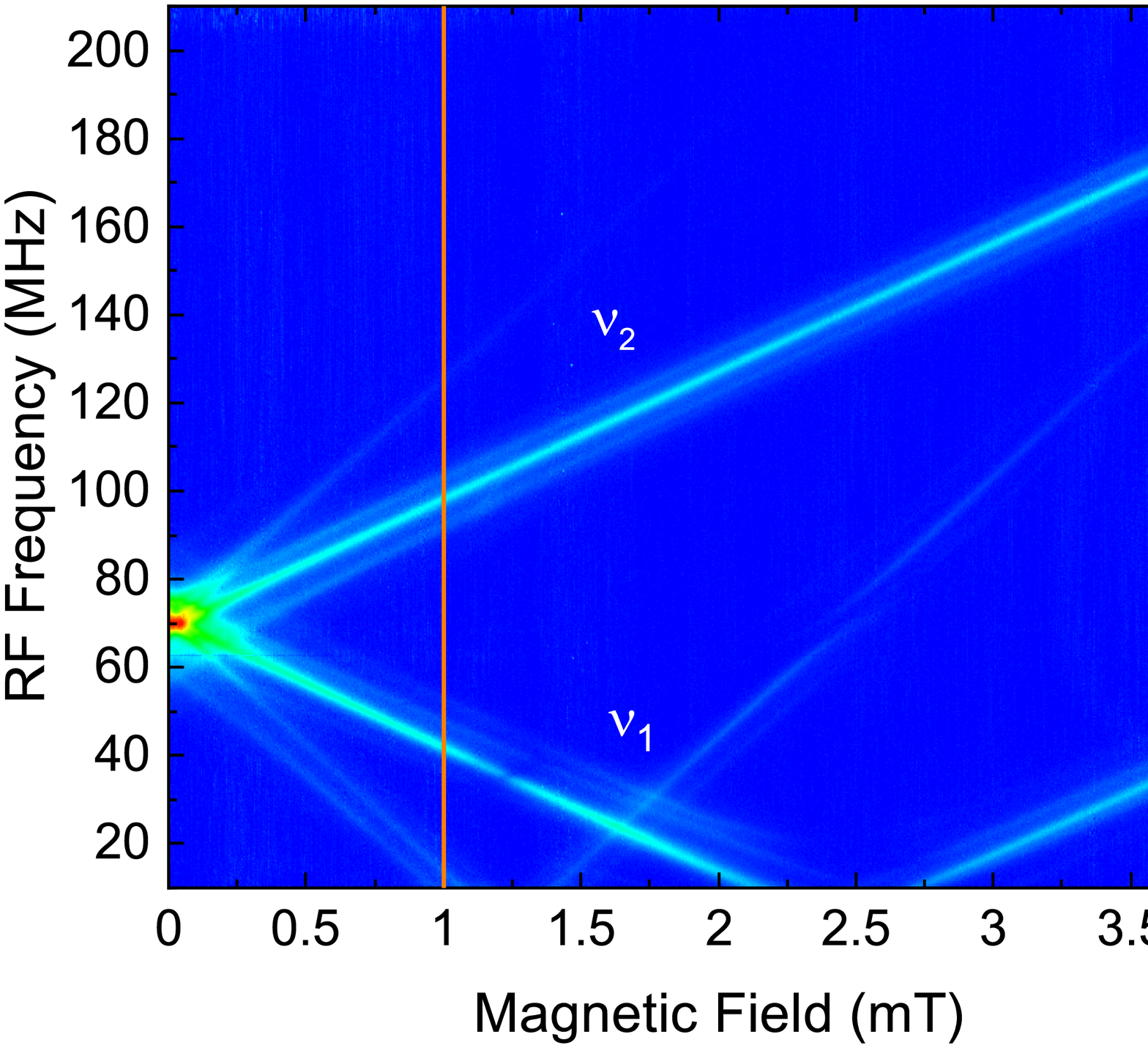}\label{odmr_1}
	\includegraphics[width=0.45\textwidth]{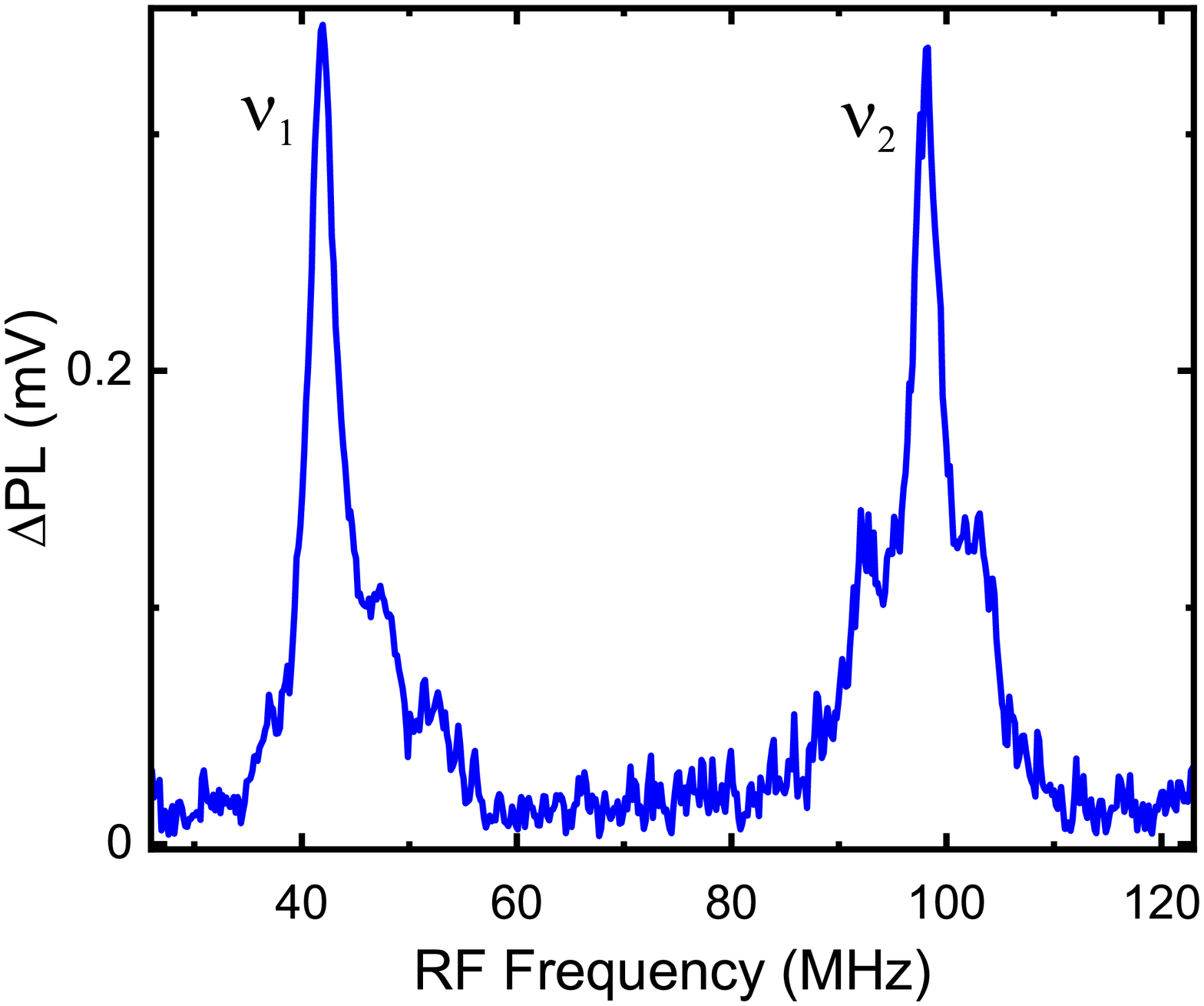}\label{odmr_2}
	\caption{(a) High resolution ODMR Spectrum for the annealed sample with 200 mW of optical power. The $\nu_1$ transition corresponds to the 1/2 $\to$ 3/2, while $\nu_2$ corresponds to the -1/2 $\to$ -3/2 transition. The other transitions are visible as faint lines. The dark -1/2 $\to$ +1/2 transition cutting through the $\nu_1$ is visible at 1.25 mT. (b) Line cut of the ODMR spectrum at 1 mT. The hyperfine coupling of $^{29}$Si observed at $\pm$ 5 MHz for the $\nu_2$ transition.}\label{odmr}
\end{figure*}

\section{Methods}

Samples are diced from high purity semi-insulating SiC substrates purchased from CREE \footnote{https://www.wolfspeed.com/materials/products/semi-insulating-sic-substrates}. The samples are neutron irradiated by a TRIGA reactor (Training, Research, and Isotope Production General Atomics) at the McClellan nuclear reactor operated by UC Davis. We irradiated samples with a fluence of  10$^{17}$ $n^0$/cm$^2$ and estimate the silicon vacancy density to be $\sim 10^{15}$ $V^{-}_{Si}$/cm$^3$ \cite{fuchs2015,kasper2020}. Post-irradiation, the samples are noticeably opaque as was described in a recent report \cite{kasper2020}. Following neutron irradiation, the samples are annealed in vacuum at a temperature of 600 C for 4 hours. A subset of samples were thermally quenched by annealing the samples in evacuated quartz ampoules, quickly removing them from the furnace at the end of the anneal and immersing them in a water bath. The quench was motivated by the simulations of Ref. \onlinecite{wang2013} which found that a thermal quench would limit the transformation of V$^{-}_{Si}$ to the carbon anti-site defect (V$_C$ C$_{Si}^{2+}$) as the sample returns to room temperature. 

The samples are subsequently studied in a custom built photoluminescence microscope with an integrated dual set of Helmholtz coils as shown schematically in Fig. \ref{exp_pl}(a). The inner coil applies an RF magnetic field perpendicular to the c-axis of the crystal to excite spin transitions. The inner coil is nested perpendicularly within a larger Helmholtz coil which applies a field parallel to the c-axis of the crystal. The larger coil applies a static magnetic field in order to split the spin transitions through the Zeeman effect as described in Eq. \ref{f_B}. A continous-wave 785-nm laser diode is focused on the sample through a 50X Nikon objective with a numerical aperture of 0.55. Based on the numerical aperture of the objective and the excitation wavelength, we estimate a luminescence volume of 0.3 $\mu$m$^3$. Multiplying by the estimated density of defects, we estimate that we are interrogating approximately 300 V$^{-}_{Si}$. For the photoluminescence study, the samples are pumped with 100 mW of power. For the ODMR measurements, the samples are optically pumped with intensities ranging from 20 to 160 mW/$\mu$m$^2$ while the outer Helmhlotz coil applies a static magnetic field of 1 mT along the c axis of the crystal. A modulated RF field is applied to the sample through the integrated Helmholtz coil along the crystal plane perpendicular to the c axis of the sample. The photoluminescence from the sample is collected by the same objective by which the sample is pumped. It is reflected off a dichroic mirror, passed through a 850 nm long pass filter, coupled into a fiber and detected by an amplified InGaAs photodiode. Based on the numerical aperture of the objective, we estimate that 11$\%$ of the emission is collected. The signal from the photodetector is output to a lock-in amplifier (Signal Recovery 7265) which detects the modulated photoluminescence produced by the resonant RF excitation.

\section{Results}

The relative photoluminescence emission for the irradiated, annealed, and annealed/quenched samples are presented in Fig. \ref{exp_pl}(b). The emission rate of the samples clearly increases with further materials processing. The photoluminescence yield of the annealed sample is a factor of 4 greater than the irradiated sample which is in reasonable agreement with a recent report \cite{kasper2020}, while the annealed and quenched sample is a factor of 6 greater, a 50$\%$ improvement upon annealing alone. This indicates that the transformation of $V^{-}_{Si}$ to V$_C$C$_{Si}^{2+}$ is potentially a limiting factor for ensemble density \cite{wang2013}.

\begin{figure}
	\includegraphics[width=0.45\textwidth]{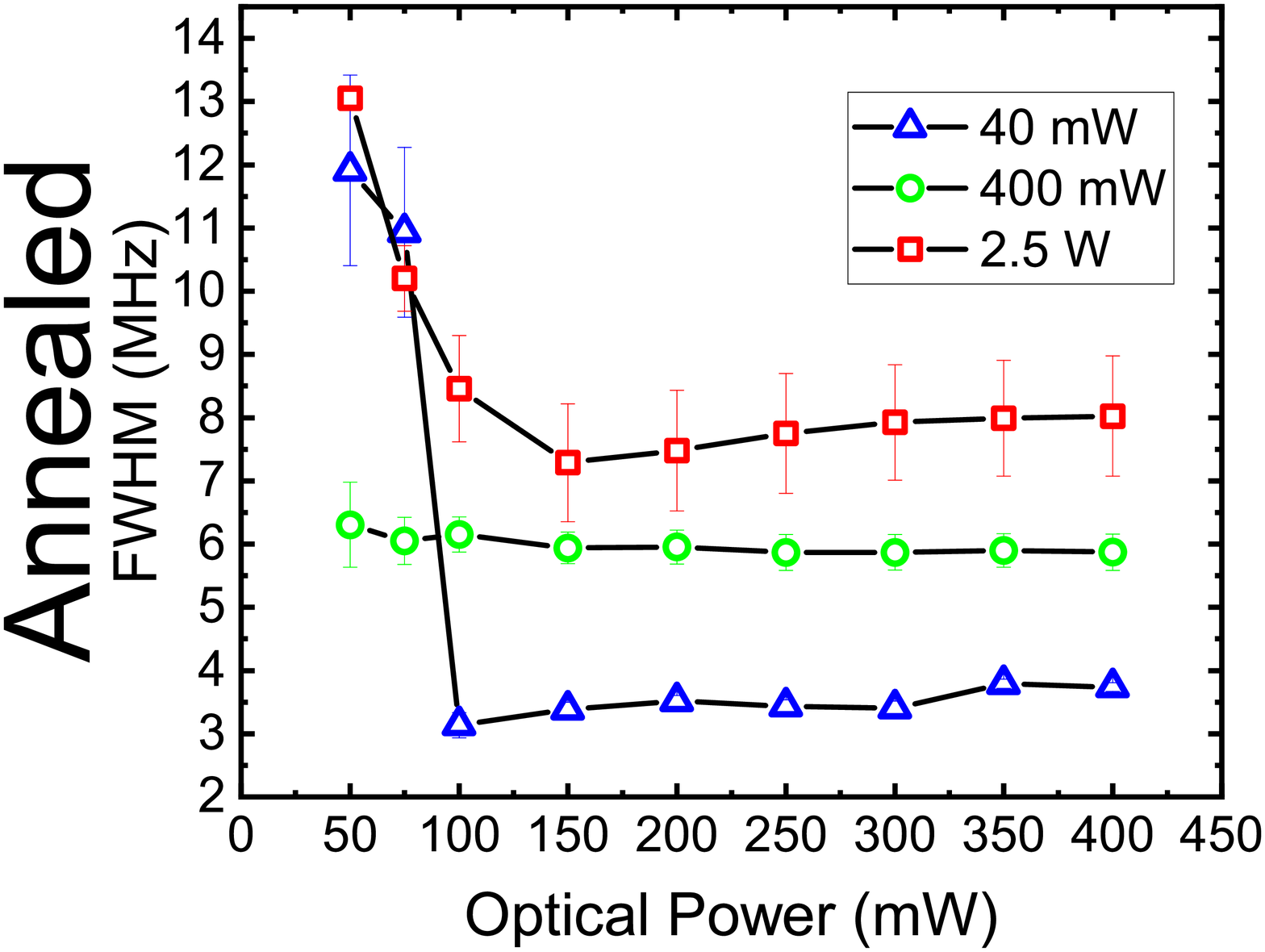}\label{fwhm_opt_a}
	\includegraphics[width=0.45\textwidth]{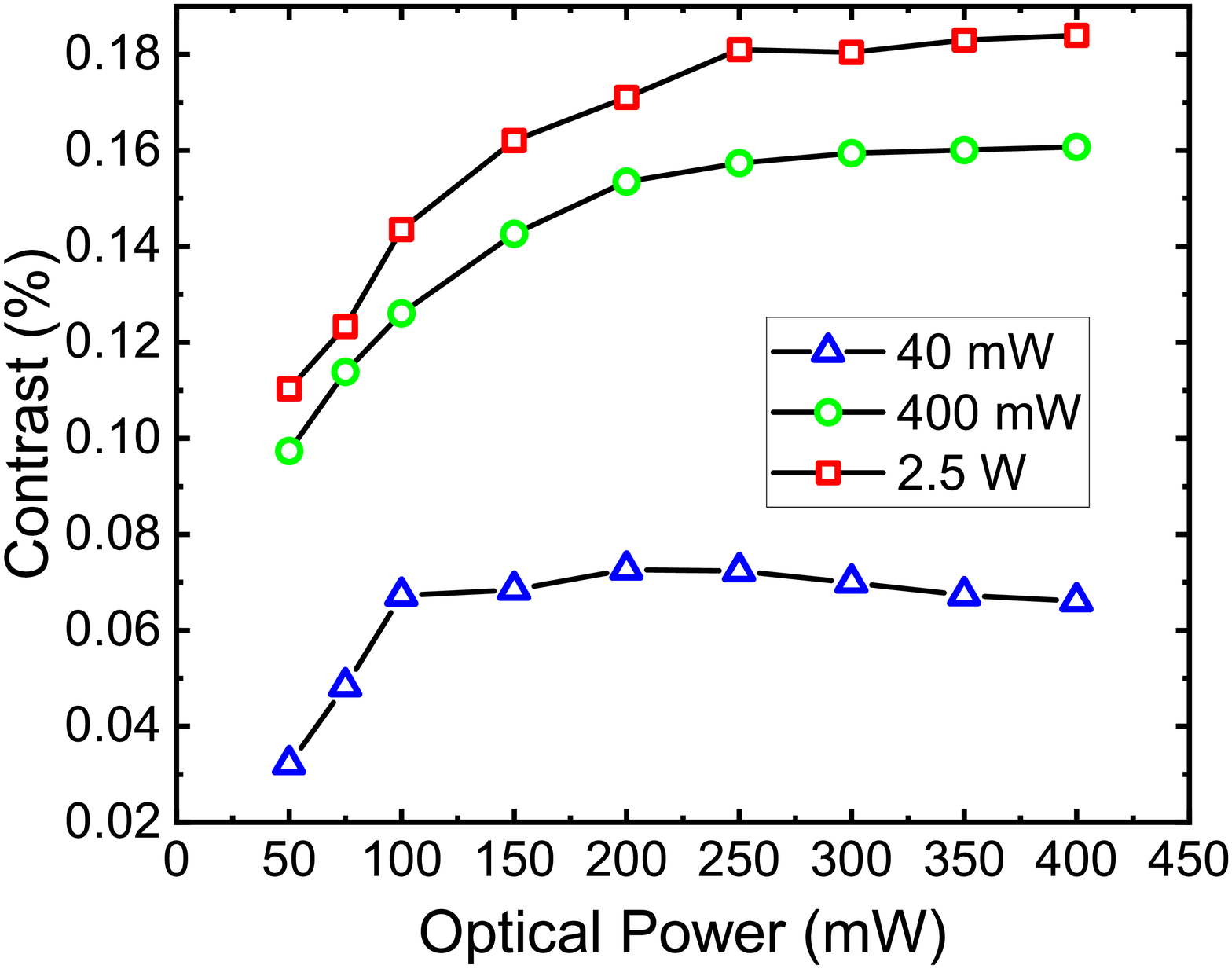}\label{c_opt_a}
	\vskip
	\baselineskip
	\includegraphics[width=0.45\textwidth]{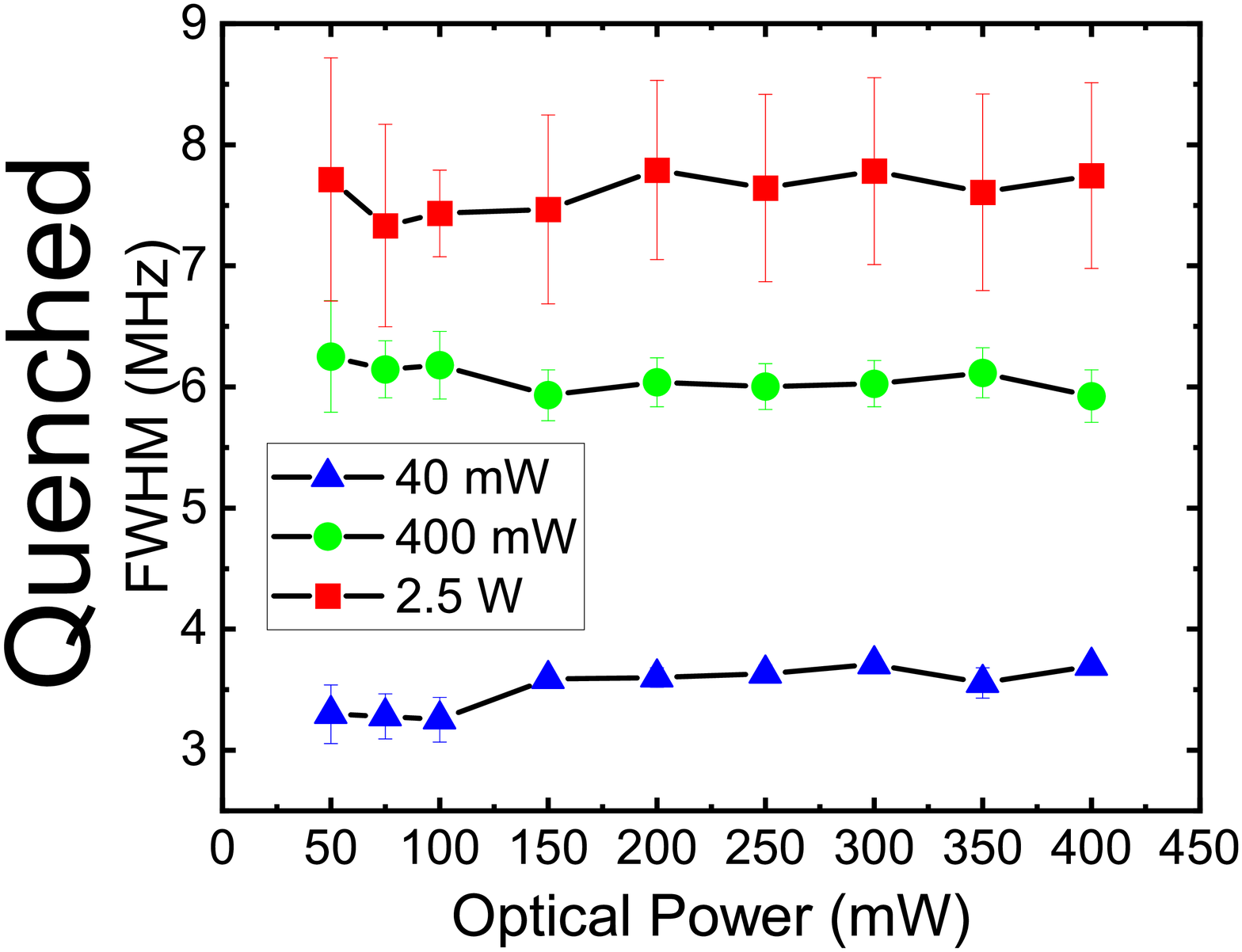}\label{fwhm_opt_q}
	\includegraphics[width=0.45\textwidth]{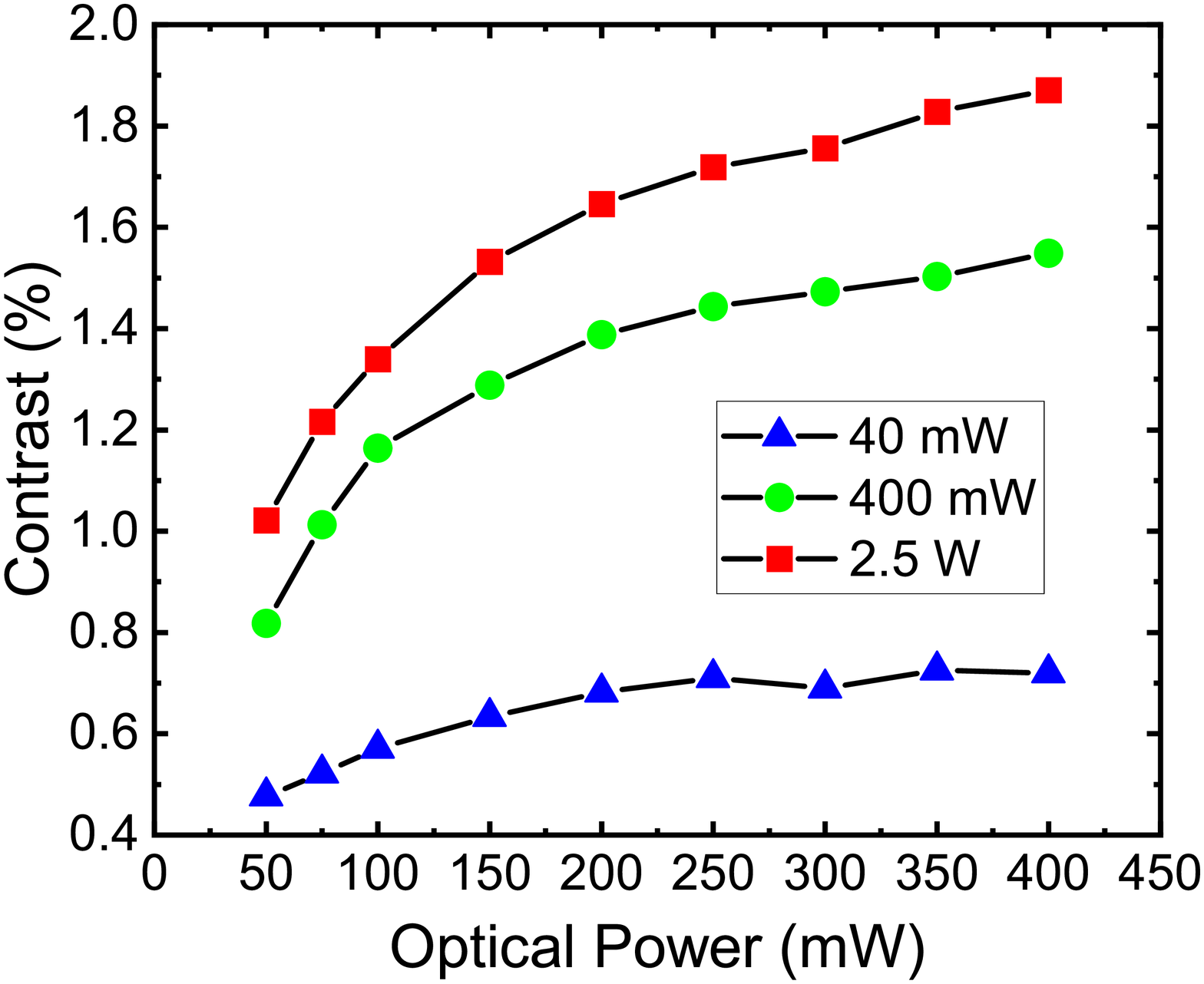}\label{c_opt_q}
	\caption{Optical power broadening measurements for the samples. From left to right, top to bottom; (a) FWHM  (b) contrast versus optical power for different RF powers with the annealed samples (c) FWHM (d) contrast versus optical power for different RF powers with the quenched sample. The error bars in (a) and (c) represent the 95$\%$ confidence intervals for the FWHM from Lorentizian fitting the data.}
\label{opt_broad}
\end{figure}

\begin{figure}
	\includegraphics[width=0.45\textwidth]{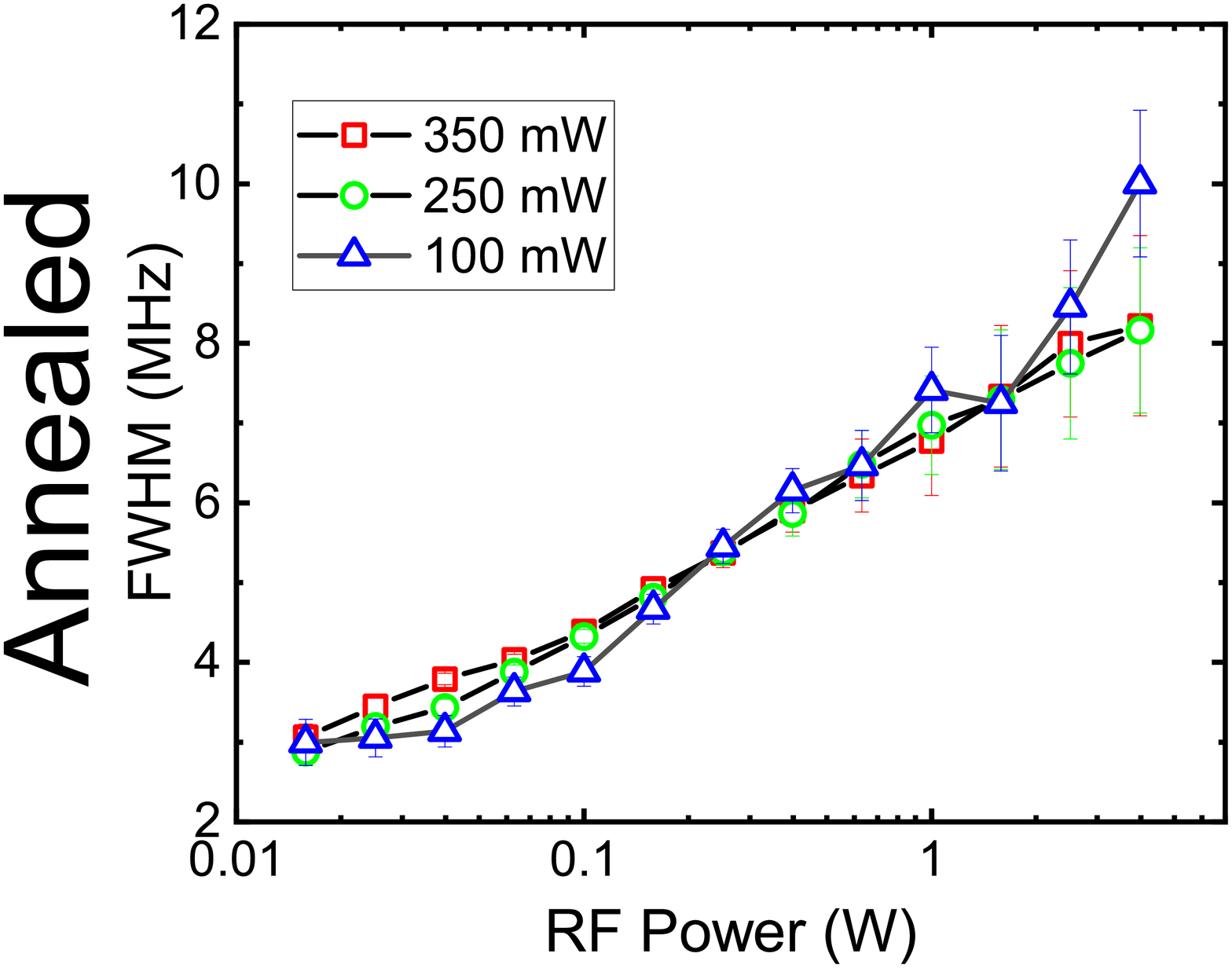}\label{fwhm_rf_a}
	\includegraphics[width=0.45\textwidth]{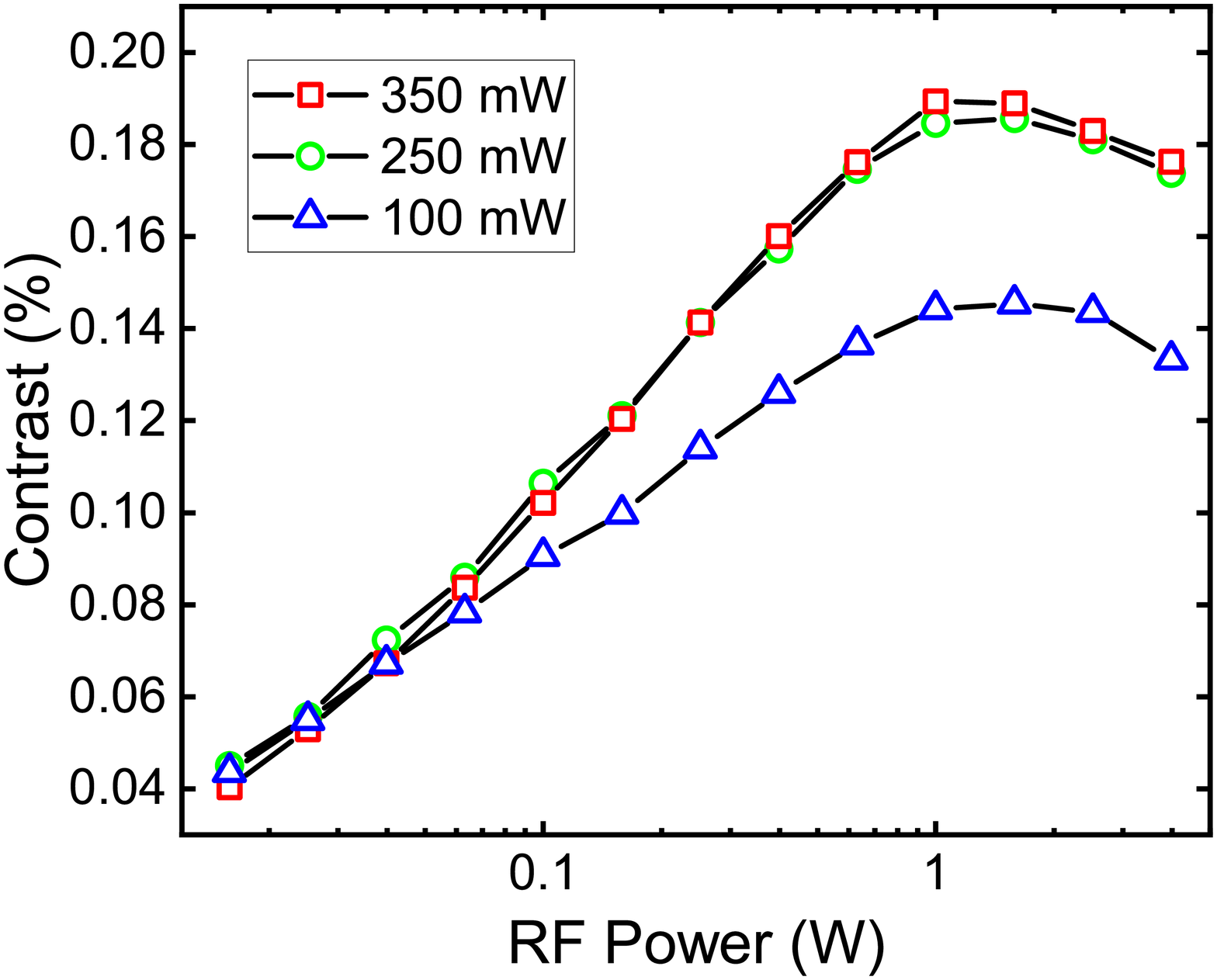}\label{c_rf_a}
	\vskip
	\baselineskip
	\includegraphics[width=0.45\textwidth]{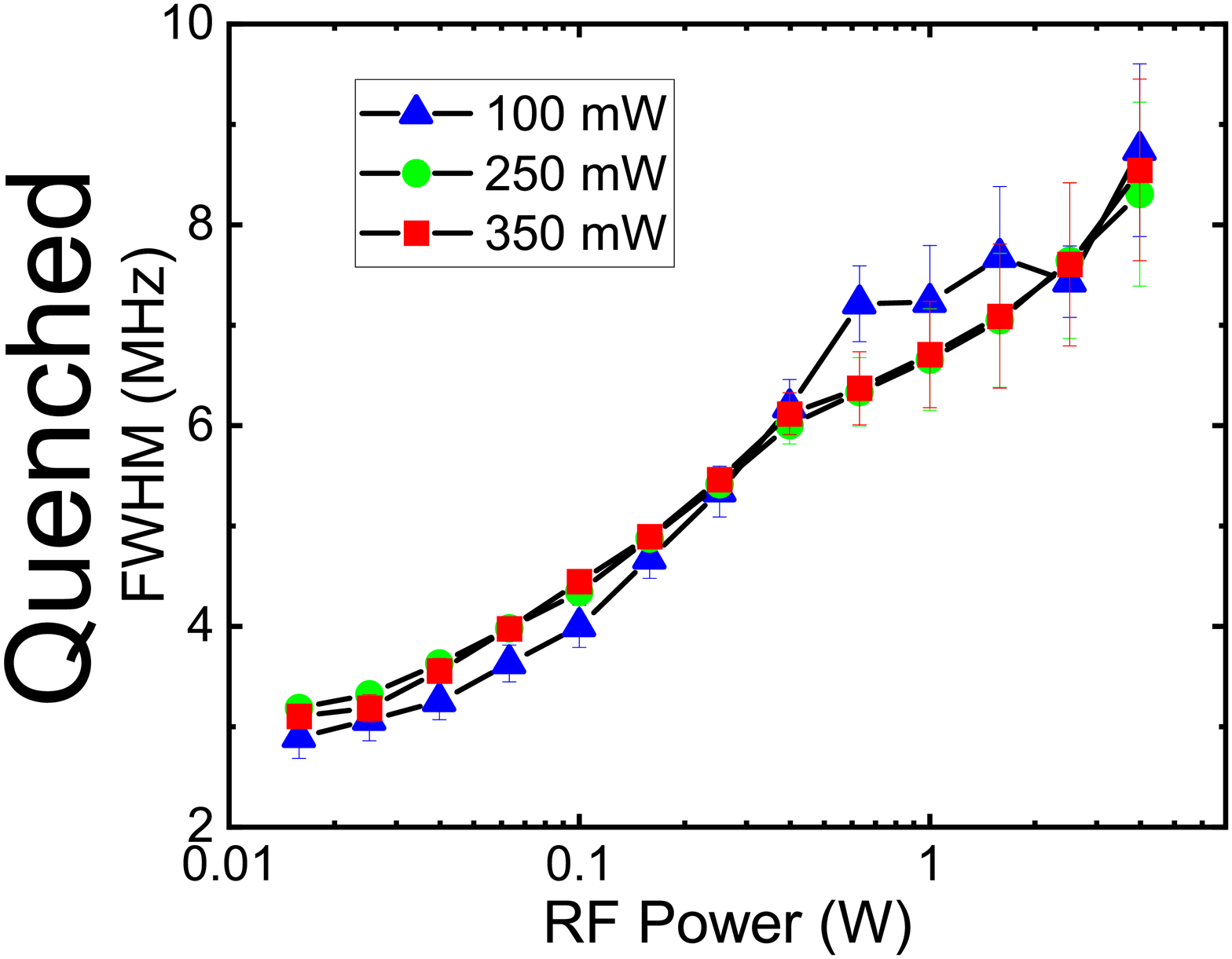}\label{fwhm_rf_q}
	\includegraphics[width=0.45\textwidth]{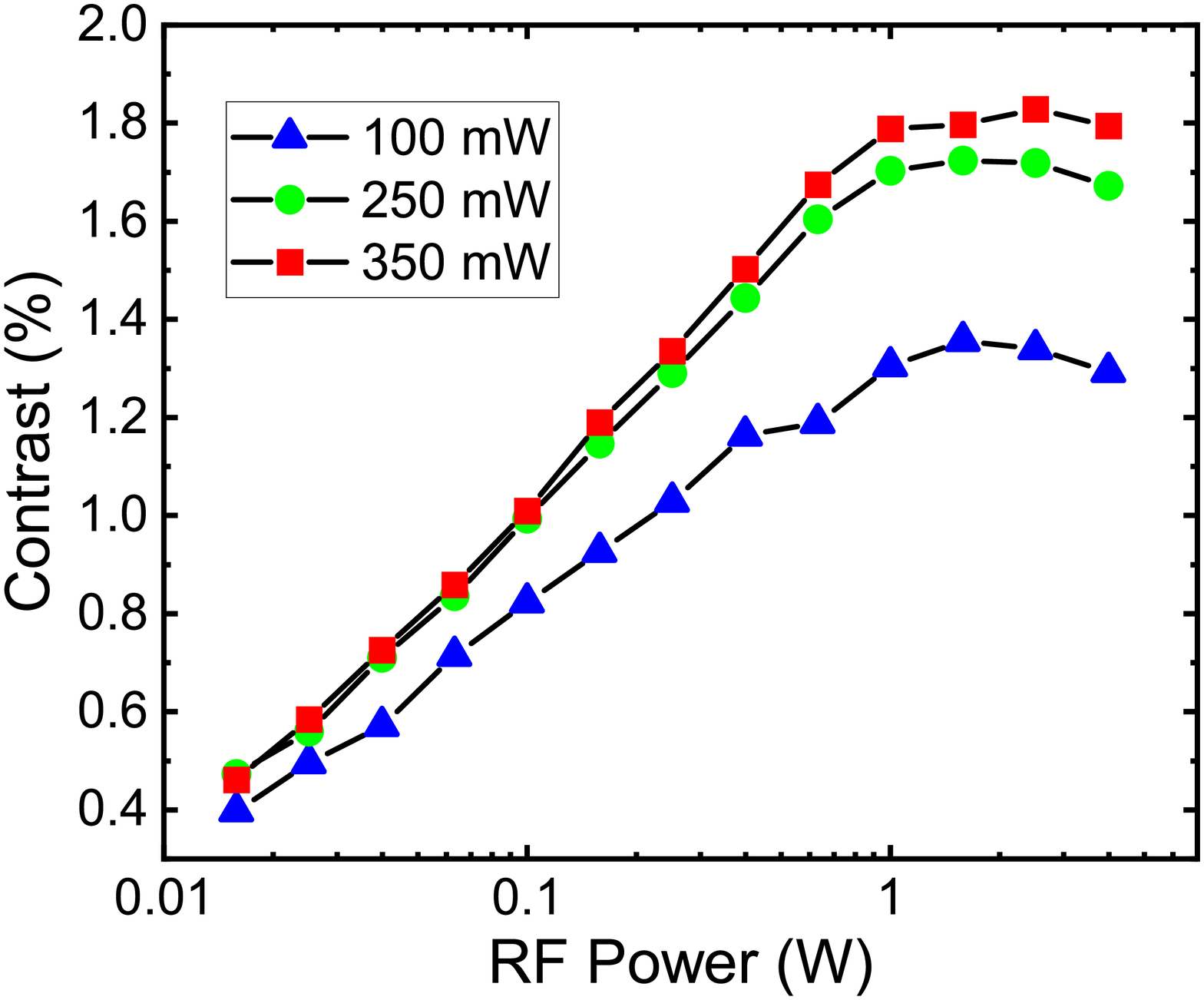}\label{c_rf_q}
	\caption{RF power broadening measurements for the samples. From left to right, top to bottom; (a) FWHM  (b) contrast versus RF power for different optical powers with the annealed samples (c) FWHM (d) contrast versus RF power for different optical powers with the quenched sample. The error bars in (a) and (c) represent the 95$\%$ confidence intervals for the FWHM from Lorentizian fitting the data.}
\label{rf_broad}
\end{figure}

After characterizing the photoluminescence emission of the samples, we performed optically detected magnetic resonance (ODMR) measurements of the samples. For these measurements, we amplitude modulate the RF drive and detect the modulated signal with a  lock-in amplifier to measure the photoluminescence change resulting from the ensemble being on resonance. Fig. \ref{odmr}(a) is a high resolution image of the ODMR spectrum of the annealed sample. The strong lines labeled $\nu_1$ and $\nu_2$ correspond to the \textit{m} = 1 transitions from +1/2 $\to$ +3/2 and -1/2 $\to$ -3/2 respectively. Faint signatures of the higher order \textit{m} = 2 transitions are visible. Additionally, the dark -1/2 $\to$ +1/2 transition can be seen intersecting the $\nu_1$ transition at 1.25 mT. The transitions are labeled according to the convention of Ref. \cite{simin2016}. The line cut in Fig. \ref{odmr}(b) corresponds to the ODMR line scan of Fig. \ref{odmr}(a) performed at bias magnetic field of 1 mT. Of note, the $^{29}$Si hyperfine coupling to the $\nu_2$ transition is observable $\approx$ 10 MHz as in Ref. \cite{kraus2014-1}. All further measurements are conducted at a bias magnetic field of 1 mT to minimize the influence of the Earth and laboratory fields.  

\begin{figure}
\centering
\includegraphics[width=1\textwidth]{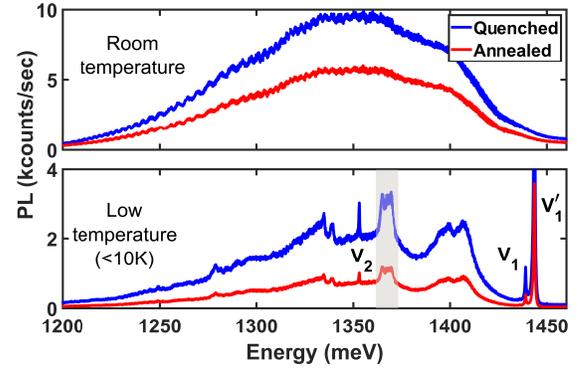}
\caption{Room temperature and low temperature photoluminescence comparison. At low temperature, the V1', V1, and V2 zero phonon lines (ZPL) are resolvable. The feature at 1370 meV is indicated with the grey shading.}\label{lt_pl}
\end{figure}

\begin{figure}
\centering
\includegraphics[width=1\textwidth]{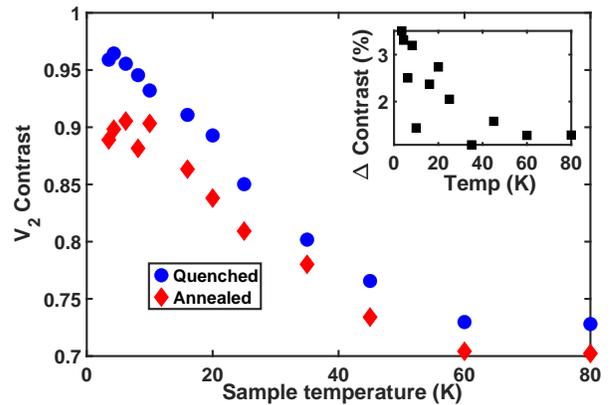}
\caption{Relative amplitude, labeled as $V_2$ Contrast, of the V2 ZPL versus the non-$V^{-}_{Si}$ peak at 1370 meV for the quenched and annealed samples versus temperature. (Inset) The relative difference of the quenched sample versus annealed sample V2 contrast.}\label{vis_v2}
\end{figure}

To study power broadening, we fix the laser and RF power and sweep the RF frequency across the $\nu_2$ transition labeled in Fig. \ref{odmr}(b). After an ODMR peak is acquired, we fit a Lorentzian peak to extract the FWHM and the height of the peak which is normalized by the voltage out of the photodetector to calculate the ODMR contrast. This process is repeated as the RF and optical powers are swept. The dependence of the contrast and FWHM as a function of RF and optical power is presented in the plots of Fig. \ref{opt_broad} and \ref{rf_broad} for both samples. The FWHM is essentially independent of optical power for both samples as seen in Fig. \ref{opt_broad} in agreement with Ref. \onlinecite{wang2020}. The primary difference in the optical broadening response between the annealed and quenched samples is that the quenched sample has a greater contrast by a factor of 10 as shown in Figs. \ref{opt_broad}(b) and \ref{opt_broad}(d). This is also observed with the RF broadening response in Figs. \ref{rf_broad}(b) and \ref{rf_broad}(d). There is general qualitative agreement between our results and Refs. \cite{jensen2013,wang2020}.

\begin{figure*}
\includegraphics[scale=0.30]{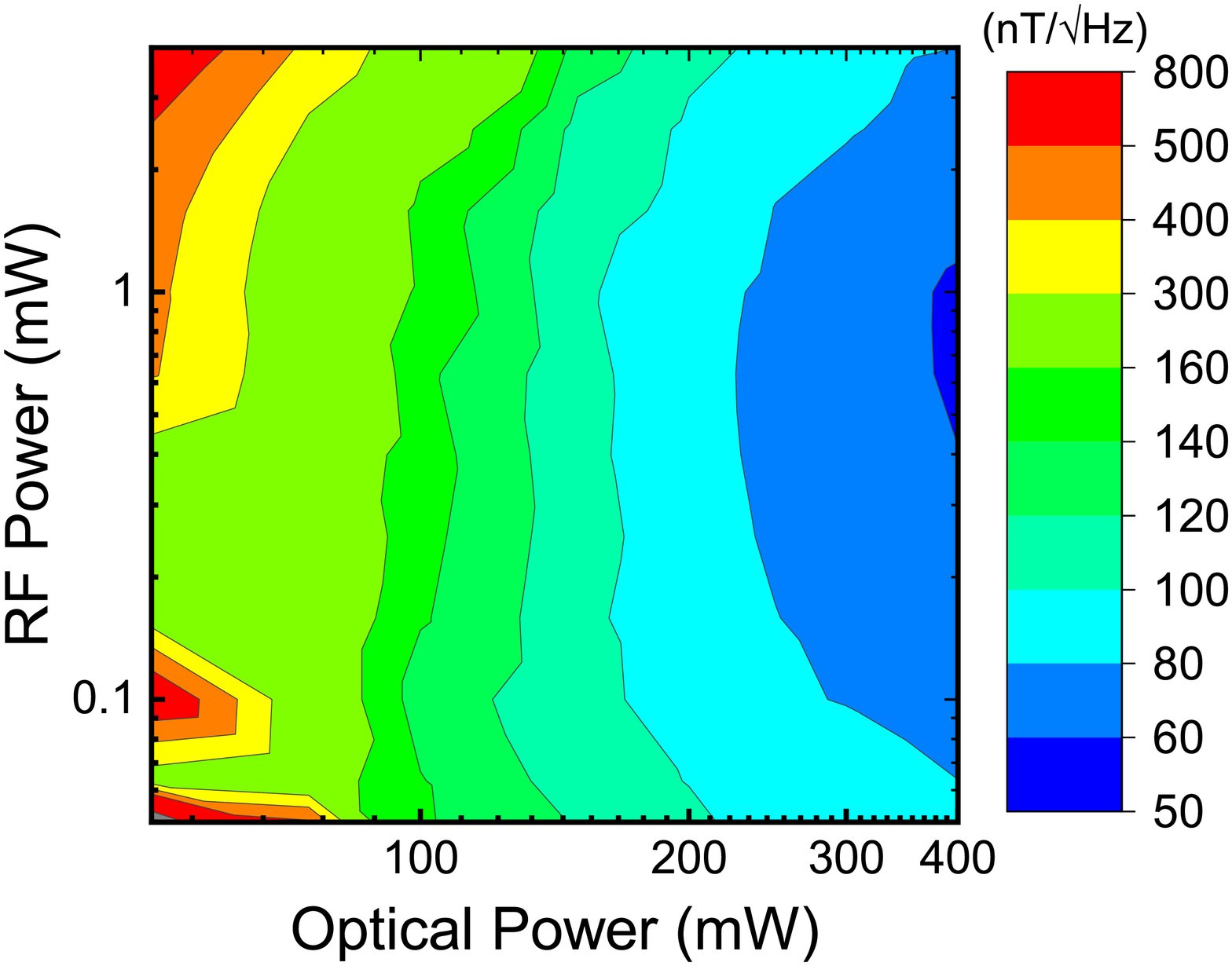}\label{a_sens}
\includegraphics[scale=0.30]{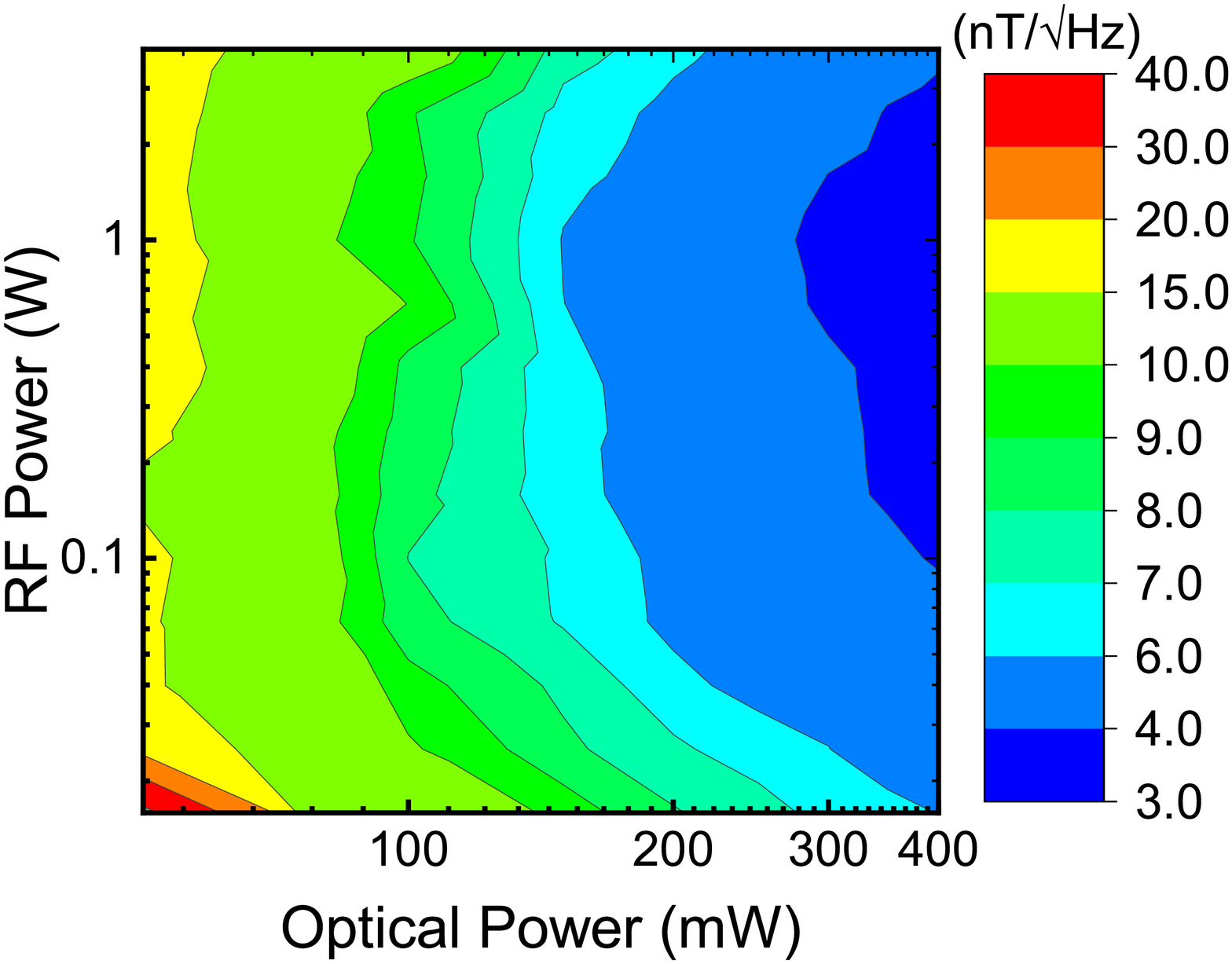}\label{q_sens}
\caption{(a) The shot noise sensitivity plot for the annealed sample. (b) The shot noise sensitivity plot for the quenched sample. The minimum value for $\eta_{B}$ is 57 and 3.5 nT/$\sqrt{Hz}$, respectively.}\label{sensitivity}
\end{figure*}

To understand the difference in contrast between the quenched and annealed samples, we measured the photoluminescence emission spectra at low temperature. The results are presented in Fig. \ref{lt_pl} and Fig. \ref{vis_v2}. The V1' ,V1, and V2 zero photon lines (ZPL) of the $V^{-}_{Si}$ are visible at 1445, 1440 and 1354 meV, respectively. The plotted intensities are scaled to display total counts/second, so this scaling results in the slight differences in the signal/noise ratio in the individual spectra. However, this scaling has no effect on the observed enhancement of V1/V2 emission and reduction of other features reported in this work. There is an unidentified feature centered at 1370 meV which is also present in the low temperature photoluminescence measurements of Ref. \cite{fuchs2015}. In Fig. \ref{vis_v2}, we compare the peak relative amplitudes of the V2 ZPL to this unidentified feature as a function of temperature since the V2 ZPL is responsible for the ODMR response \cite{castelletto2020}. The peak relative amplitude of the V2 ZPL to the 1370 meV feature for the quenched sample is consistently larger than with the annealed sample as temperature is lowered. Additionally, the peak relative amplitude for the quenched sample increases as temperature is lowered. Extrapolating from the results from the anneal kinetics of Ref. \cite{wang2013}, that the thermal quench reduces the transformation of the $V^{-}_{Si}$ to other defects, we can infer that the quench also minimized the formation of the defect responsible for the 1370 meV peak. With the working model that the improvement in materials processing through quenching the sample is driving the improved brightness and contrast of the quenched sample, it is likely that the improved ODMR contrast of the quenched sample is due to two synergistic factors: the relative increase in the formation of the silicon vacancy and the relative decrease in the formation of the unknown bright defect at 1370 meV which has spectral overlap with the $V^{-}_{Si}$. 

\begin{figure}
\centering
\includegraphics[width=1\textwidth]{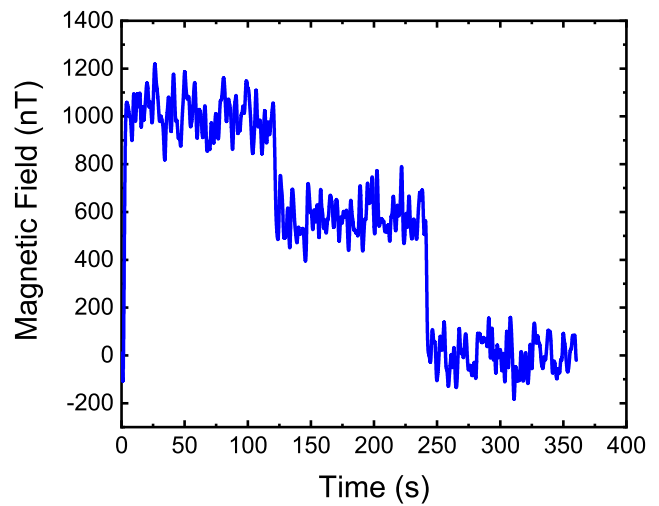}
\caption{Magnetic field step response. The magnetic field is stepped by 500 nT every 120 seconds. For this measurement, the lock-in time constant is set to 0.5 seconds. The standard deviation of the steps is 70 nT. This results in a measured sensitivity of 50 nT$/\sqrt{Hz}$.}\label{noise}
\end{figure}

Lastly, we optimized the applied RF and optical powers to minimize the shot noise sensitivity. This is a critical step to accurately benchmark the ultimate sensitivity of a quantum defect ODMR-based magnetometer \cite{dreau2011,shin2012,jensen2013,clevenson2015,clevenson2018,budker2017,wang2020}. To find the optimal RF and optical power, we calculate the shot noise sensitivity according to Eq. \ref{sn_sensitivity} by measuring the FWHM, contrast, and photon emission rate of the ODMR response. The FWHM is measured by fitting a Lorentzian peak to the ODMR response. The contrast is calculated by taking the lock-in voltage at the ODMR peak, subtracting the lock-in baseline voltage, and dividing the remainder by the DC voltage of the photodetector. We measure the photon emission rate by converting the voltage out of the detector to a rate of photons through the responsivity and trans-impedance gain of the detector. In Fig. \ref{sensitivity}, plots of the sensitivity maps for the annealed Fig. \ref{sensitivity}(a) and the quenched Fig. \ref{sensitivity}(b) samples are presented. The quenched sample has a factor of 15 improved sensitivity relative to the annealed sample while the improvement of the quenched versus annealed photoluminescence intensity is only a factor of 1.5 higher as shown in Fig. \ref{exp_pl}. From Eq. \ref{sn_sensitivity}, we would expect an improvement in the shot noise sensitivity of a factor of $\frac{1}{10\sqrt{1.5}} \approx 0.08$ from the increase in the $V^{-}_{Si}$ PL from quenching. Thus, the improved sensitivity is primarily due to the higher contrast of the quenched sample relative to the annealed samples shown in Fig. \ref{opt_broad}(b) and Fig. \ref{opt_broad}(d). Based on the measured parameters, we compute the shot-noise-limited sensitivity to be 3.5 nT$/\sqrt{Hz}$, which corresponds to the quenched sample with an optical power of 400 mW and an RF power of 1 W. We anticipate that the sensitivity will eventually improve with increasing the optical power as has been previously observed \cite{wang2020}. In our experimental system, the optical power is limited to 400 mW.

To measure the actual noise of the magnetometer, we apply a frequency modulated RF tone to the inner Helmholtz coils, see Fig. \ref{exp_pl}(a). The signal from the lock-in is proportional to the frequency offset from resonance. The lock-in time constant is set to 0.5 seconds. We then step the applied DC magnetic field by 500 nT steps every 120 seconds. The result of the measurement is presented in Fig. \ref{noise}. The standard deviation of each step is 70 nT. Scaling by the measurement frequency, the measured noise of the magnetometer is 50 nT$/\sqrt{Hz}$. This is a factor of 2 improvement over a recently reported measurement with the silicon vacancy in silicon carbide \cite{simin2016} and a factor of 167 greater than a recently reported measurements with the DNV \cite{patel2020}.

\section{Conclusion}

We have demonstrated magnetometry with the $V^{-}_{Si}$ in 4H-SiC with a measured sensitivity of 50 nT$/\sqrt{Hz}$ and a theoretical shot-noise-limited sensitivity of 3.5 nT/$\sqrt{Hz}$. This is the most sensitive demonstration of magnetometry with this defect to date.  We accomplish this by improving the material properties of the sample through rapidly quenching the sample post-annealing, and by characterizing the power broadening of the system to optimize the optical and RF powers. Our low temperature PL measurements indicated that the improved contrast and sensitivity of the quenched sample may be related to an unidentified defect with an emission line at 1370 meV. To further characterize the relative density of this competing defect, double electron-electron resonance measurements of the samples could provide further insight. 

Our result opens a number of new avenues for improving the yield of silicon vacancies in silicon carbide and performing magnetometry with ensembles of the defect. Based on the improvements of defect density and contrast with quenching the sample after annealing  and previous results with the laser formation of the $V^{-}_{Si}$ \cite{chen2019,wang2019}, ultra fast laser processing \cite{phillips2015} may be a viable path to further improve the yield of defect formation for neutron irradiated samples since it would dramatically reduce the timescale of defect formation and the return to equilibrium. Additionally, greater sensitivity may be attainable through annealing and quenching electron irradiated samples. This is of interest because of the recent report \cite{kasper2020} of greater T$_{2}$ times observed with electron versus neutron irradiation. The coarse optimization of anneal parameters can be expanded upon to understand if greater improvements are possible. Ref. \cite{wang2013} performed their simulation for the 4H polytype. Expanding their work by investigating the relative energies of the $V^{-}_{Si}$ and V$_C$C$_{Si}$ $^{2+}$ in other polymorphs such as SiC 3C could point to another improvement path. In addition to materials improvement, we can expect improved sensitivity through implementing pulsed ODMR \cite{budker2017,barry2020} and other coherent techniques. 

This work demonstrates that although the $V^{-}_{Si}$ is not as bright or high in ODMR contrast as the DNV \cite{budker2017, simin2016}, it can still be developed into a sensitive magnetometer. This is a critical development given the advantages of SiC over diamond with regards to cost and mature microfabrication techniques. Additionally, electroluminescence \cite{lohrmann2015} has been demonstrated with the $V^{-}_{Si}$ whereas it has been attempted repeatedly with the DNV with no success. Thus, one can imagine a quantum magnetometer integrated on chip with silicon carbide without the need for optical integration with a sensitivity comparable to the DNV. 

\begin{acknowledgements}
This work is supported by the Johns Hopkins University Applied Physics Laboratory Project Innovation. LP and TMM were supported by the David and Lucile Packard Foundation. The annealing work was supported as part of the Institute for Quantum Matter, an Energy Frontier Research Center funded by the U.S. Department of Energy, Office of Science, Office of Basic Energy  Sciences, under Award No. DE-SC0019331. We thank Sophia Economou for helpful conversation, Tomek Kott for reading and commenting on the manuscript. We would also like to thank Samuel Carter and Evan Glaser for lending us a sample when we began our work.
\end{acknowledgements}

\bibliography{ms}

\end{document}